\documentclass[aps, prb, reprint, superscriptaddress]{revtex4-1}

\usepackage{xcolor}
\usepackage{amsmath}
\usepackage{graphicx}

\newcommand{\aloxal}{Al--AlO$_\mathrm{x}$--Al}
\newcommand{\alox}{AlO$_\mathrm{x}$}
\newcommand{\corundum}{Al$_2$O$_3$}
\newcommand{\ket}[1]{\left|#1\right>}
\newcommand{\bra}[1]{\left<#1\right|}
\newcommand{\B}[1]{\mathbf{#1}}

\begin{document}

\title{The effect of atomic structure on the electrical response of \\ aluminium oxide tunnel junctions}
\date{\today}

\author{M. J. Cyster}
\affiliation{Chemical and Quantum Physics, School of Science, RMIT University, Melbourne, Australia}

\author{J. S. Smith}
\affiliation{Chemical and Quantum Physics, School of Science, RMIT University, Melbourne, Australia}

\author{J. A. Vaitkus}
\affiliation{Chemical and Quantum Physics, School of Science, RMIT University, Melbourne, Australia}

\author{N. Vogt}
\affiliation{Chemical and Quantum Physics, School of Science, RMIT University, Melbourne, Australia}
\affiliation{HQS Quantum Simulations GmbH, 76187 Karlsruhe, Germany}

\author{S. P. Russo}
\affiliation{Chemical and Quantum Physics, School of Science, RMIT University, Melbourne, Australia}

\author{J. H. Cole}
\email[Email: ]{jared.cole@rmit.edu.au}
\affiliation{Chemical and Quantum Physics, School of Science, RMIT University, Melbourne, Australia}

\begin{abstract}
    Many nanoelectronic devices rely on thin dielectric barriers through which electrons tunnel.
    For instance, aluminium oxide barriers are used as Josephson junctions in superconducting electronics.
    The reproducibility and drift of circuit parameters in these junctions are affected by the uniformity, morphology, and composition of the oxide barriers.
    To improve these circuits the effect of the atomic structure on the electrical response of aluminium oxide barriers must be understood.
    We create three-dimensional atomistic models of aluminium oxide tunnel junctions and simulate their electronic transport properties with the non-equilibrium Green's function formalism.
    Increasing the oxide density is found to produce an exponential increase in the junction resistance.
    In highly oxygen-deficient junctions we observe metallic channels which decrease the resistance significantly.
    Computing the charge and current density within the junction shows how variation in the local potential landscape can create channels which dominate conduction.
    An atomistic approach provides a better understanding of these transport processes and guides the design of junctions for nanoelectronics applications.
\end{abstract}

\maketitle

\section{Introduction}

Superconducting qubits are one of the most promising architectures for quantum computers and are currently the favoured technology for many quantum computing groups around the world. \cite{Clarke2008,Devoret2013,Kandala2017,Mohseni2017,Nersisyan2019}
These qubits rely on the non-linear response of Josephson junctions which are often fabricated as \aloxal\ tri-layer junctions. \cite{Guangming2013,Satoh2015,Hutchings2017}
As we move to large-scale quantum computer engineering, it becomes critical to understand what limits junction performance and variability.

The conductance of \aloxal\ junctions is commonly understood in a simplified one-dimensional picture.
In the simplest case, the tunnel junction is considered to be a rectangular barrier where the transmission probability of an incident electron can be calculated using the WKB equations. \cite{Hartman1964}
More detailed analytic models of the tunnelling barrier include corrections for temperature, applied voltage, image forces, and asymmetries. \cite{Stratton1962}

Two of these models -- the Simmons model,\cite{Simmons1963} and the Brinkman, Dynes and Rowell model\cite{Brinkman1970} -- are often used to estimate parameters such as the barrier height and the oxide thickness by fitting to experimental measurements. \cite{Morohashi1987,Barner1989,Dorneles2003,Holmqvist2008,Jung2009,Aref2014}
Barrier heights calculated by fitting to the Simmons model\cite{Dorneles2003,Holmqvist2008,Jung2009} range from 0.8~eV to 3.0~eV while a ``typical'' height of 2~eV is often quoted.\cite{Snow1996,Jeurgens2002,Gloos2003,Hasnaoui2005}
Estimates of the barrier height and oxide thickness given by such models are effective values which include contributions from oxide properties such as the density and stoichiometry implicitly.
While useful, one-dimensional descriptions of the barrier system are unable to fully represent the amorphous oxide layer.

To include the full three-dimensional structure of the junction we turn to a numerical approach.
There is a growing body of literature in which the non-equilibrium Green's function (NEGF) formalism is used to calculate the electronic properties of nanoscale devices.
This is a numerical method which allows us to calculate properties such as the transmission probability, current, and charge density.
A range of systems have been studied with this approach including graphene, silicon and phosporus-in-silicon nanowires, and carbon nanotubes.\cite{Luisier2006,Waldron2006,Svizhenko2007,Koswatta2007,Yazyev2010,Smith2015}
These systems all consist of regular repeating units which allow for the calculation to be performed in reciprocal space, potentially yielding improvements in computational efficiency.
However the \aloxal\ junctions which are the subject of this study are inherently disordered; this removes any symmetries we might exploit to reduce the complexity of the problem.

The computational challenges which arise when dealing with disordered systems may explain the small number of first-principles calculations in the literature with a focus on \aloxal\ junctions.
One study by \citet{ZemanovaDieskova2013} presents \textsl{ab-initio} transport calculations for small atomistic junction models.
The conductance was calculated using a transfer matrix method and compared to the conductance of rectangular and trapezoidal barriers as well as an \textsl{sp}-like tight-binding model.
A ground-state \textsl{ab-initio} simulation is used to determine the parameters of the tight binding calculation.
Relatively poor agreement with experimentally reported conductances is observed.
Inaccurate estimation of the barrier thickness with the Simmons model is raised as a possible cause for this discrepancy.

In this paper we use molecular dynamics techniques to create three-dimensional models of \aloxal\ junctions that include the detail of the atomic structure.
The shape of the potential barrier -- used as an input to our electronic transport model -- is calculated in three dimensions from the atomic positions and charges.
The electronic properties of the junction models are calculated with the NEGF formalism.\cite{Datta2005}
Due to the native disorder in the oxide noted above we calculate solutions to the NEGF equations for a three-dimensional real space representation of the system.
By starting with a model of the atomic structure of the \aloxal\ junction and retaining a full three-dimensional description of the structure through each part of our calculations, we probe the effect of structural changes on commonly measured quantities such as the junction resistance.

The structure of the present work is as follows.
In Sec.~\ref{sec:junction_model} we describe our approach to creating atomistic models of \aloxal\ junctions.
The non-equilibrium Green's function formalism used to calculate the electronic properties of the junction models is laid out in Sec.~\ref{sec:transport_model}.
The results presented in Sec.~\ref{sec:results} demonstrate the way in which changes in the material properties of the oxide layer such as thickness, stoichiometry, and density affect the junction resistance.
Variation in the local structure at the Al/AlO$_\mathrm{x}$ interfaces is shown to affect the uniformity of current flow through the junction.

\section{Atomistic junction model}
\label{sec:junction_model}

\begin{figure}
    \includegraphics[width=1\columnwidth]{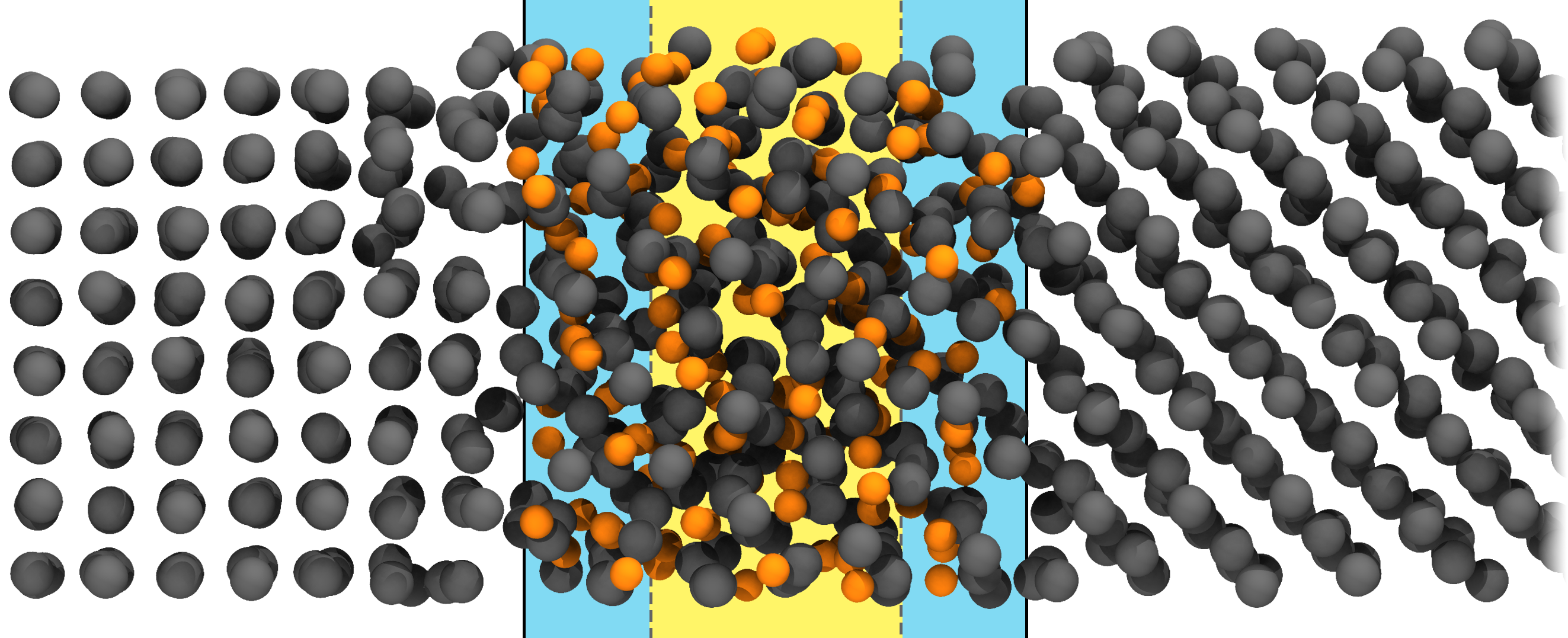}
    \caption{
        An atomistic model of a Josephson junction ($d = 14~$\AA, $\rho = 0.7$, $\gamma = 1.1$) created using a simulated annealing method.
        Aluminium and oxygen atoms are shown as grey and orange spheres respectively.
        The yellow and blue regions correspond to the central and interfacial parts of the oxide barrier respectively which are referenced in Sec.~\ref{sec:rna} and Fig.~\ref{fig:resistances_stoichiometry}b.
    }
    \label{fig:example_junction}
\end{figure}

To study the effect of atomic scale structure on the electronic properties of the junction we create atomistic models, starting with a large supercell of crystalline \corundum\ (corundum).
We adopt a convention where the thickness of the barrier $d$ is measured along the $z$-axis while $x$ and $y$ are the lateral directions.
The variable $\rho$ is used to describe the density of the junction oxide as a multiple of the density of crystalline \corundum ~(3.97~g~cm$^{-3}$).\cite{CRC} 
The variable $\gamma$ represents the stoichiometric ratio of oxygen to aluminium in the centre of the oxide (the yellow region in Fig.~\ref{fig:example_junction}).
A crystalline \corundum\ structure would therefore be described by values of $\rho = 1.0$ and $\gamma = 1.5$.

By modifying the \corundum\ crystal we produce oxide structures with a range of densities and stoichiometries.
Bulk amorphous \alox\ is experimentally reported to have lower density and stoichiometry than the crystalline structure.\cite{Sullivan1998a}
To create an oxide barrier of a given thickness $d$ but a reduced density, a volume is cut from the corundum supercell of size $\Delta x \times \Delta y \times \rho d$ after which the structure is expanded in the $z$-direction by a factor of $\rho^{-1}$.
The desired stoichiometry is then obtained by randomly removing oxygen atoms from the structure.
Following this, a geometry optimisation is performed to find the lowest energy configuration of the atoms during which the atoms are free to move, but the size of the simulation box is fixed to ensure that the density remains constant.
We use the General Utility Lattice Program (GULP) for both this optimisation and the subsequent molecular dynamics calculations.\cite{Gale1997}
Interations between the aluminium and oxygen atoms are described with an empirical potential parameterised by Streitz and Mintmire.\cite{Streitz1994}

To introduce disorder in the structure we run a molecular dynamics calculation at 3300~K (which is 1000~K above the melting point of corundum) for 4~ps with a time step of 1~fs.
Following this the simulation temperature is linearly reduced to 300~K over 6~ps to quench the oxide in a specific disorder configuration.
Crystalline aluminium regions are then placed adjacent to the oxide (in the positive and negative $z$-directions) and a second geometry optimisation is performed to reconstruct the interfacial regions between the oxide and the aluminium contacts.\cite{DuBois2015b}
During this optimisation the box can expand or contract along the $z$-axis, and atoms in the aluminium contacts and up to 4~\AA\ into the oxide on each side are free to move.
By fixing the atoms in the central region we are able to retain the desired density and stoichiometry even in cases where the final structure may not be energetically optimal.
An example of an atomistic junction model produced in this way is shown in Fig.~\ref{fig:example_junction}.

Experimental studies of aluminium oxide structure\cite{Tan2005,Sullivan1998a} show that the amorphous phase has a density approximately 0.8 times that of the crystal phase and a stoichiometry of $\gamma = 1.10$.
The oxide layer in a single junction\cite{Zeng2015a} varies between 10 and 20~\AA.
On the basis of these values we create three sets of junction models summarised in Table~\ref{tab:datasets}.
In each data set we vary one parameter while keeping the other two fixed at realistic values based on experimental data.

\begin{table}
    \caption{
        Three sets of junction models are constructed in response to experimentally reported values of the barrier thickness, the oxide density, and the oxide stoichiometry.\cite{Tan2005,Sullivan1998a,Zeng2015a}
    }
    \vspace{6pt}
    \begin{tabular}{c|c|c|c}
    Data set & Thickness (\AA) & Density ($\rho$) & Stoichiometry ($\gamma$) \\ \hline
    1 & 10--30 & 0.8 & 1.1 \\
    2 & 14 & 0.6--1.0 & 1.1 \\
    3 & 14 & 0.8 & 0.3--1.5 \\
    \end{tabular}
    \label{tab:datasets}
\end{table}

The lateral dimensions of each junction model are $\Delta x = \Delta y = 24$~\AA.
By comparison, the size of junctions in real circuits usually exceeds 100~nm.\cite{Aref2014,Satoh2015}
For this reason periodic boundary conditions are applied during the development of the junction model.

\section{Electronic transport model}
\label{sec:transport_model}

In order to calculate the electronic transport properties of the atomistic \aloxal\ junction models we implement a non-equilibrium Green's function model in one, two, and three dimensions.
This provides a single particle description of a free electron moving through a disordered potential obtained from the atomistic model.
A finite difference approximation of the kinetic operator is used to describe the channel through which transmission occurs and the source and drain contacts which are connected to it.
In the three-dimensional transport calculation the boundary conditions are periodic in $x$ and $y$, with open boundary conditions in $z$.

To second-order the finite difference representation of the kinetic energy for a one-, two-, or three-dimensional system can be written as
\begin{equation}
    T = \sum_{i}^{N} \varepsilon \ket{i}\bra{i} - \sum_{\left<i, j\right>}^{N} t_{k} \ket{i}\bra{j}
\end{equation}
where $k \in \{x, y, z\}$. 
The magnitude of the on-site energy $\varepsilon$ changes with dimensionality:
\begin{equation}
    \varepsilon =  
    \begin{cases}
        2 t_z,    &   \mathrm{1D} \\
        2 t_z + 2 t_y,    &   \mathrm{2D} \\
        2 t_z + 2 t_y + 2 t_x,    &   \mathrm{3D}
    \end{cases} 
\end{equation}
The magnitudes of the hopping energies $t_k~=~\hbar^2 / 2 m^{*} a_k^2$ are determined by the spacing between points $a_k~\simeq~1 / 3$~\AA\ and the effective mass $m^{*}$.
We choose $m^{*}$ to be the free electron mass $m_e$ as the model is designed to describe electrons tunnelling between two contacts composed of bulk aluminium in which $m^{*} \simeq m_e$.\cite{Ashcroft1965}

The electrostatic potential $V(x,y,z)$ in the junction structure is calculated on a Cartesian grid and added to the kinetic energy $T$ to form the complete channel Hamiltonian $H_C = T + V$.
Details of the numerical approximations made when computing the electrostatic potential are given in Appendix \ref{sec:potential}.
To obtain the transmission function $T(E)$ we calculate the retarded Green's function
\begin{align}
    G^{r}(E) &= \left[(E + i\eta)I - H \right]^{-1} \\
             &= \left[(E + i\eta)I - H_C - \Sigma_S(E) - \Sigma_D(E) \right]^{-1} \label{eqn:green_function}
\end{align}
where $I$ is the identity matrix, $i\eta$ is a positive imaginary infinitesimal number, and $\Sigma_{S}(E)$ and $\Sigma_{D}(E)$ are the self energies for each contact where the subscripts $S$ and $D$ denote the source and drain respectively.
The matrix inversion in Eqn.~\ref{eqn:green_function} is computationally expensive and we take advantage of a recursive algorithm to speed up our calculations.\cite{Cauley2011}

 The trace over the product of the retarded Green's function and the broadening matrices $\Gamma_{S,D} = i (\Sigma_{S,D} - \Sigma_{S,D}^{\dagger})$ yields the probability of transmission through the channel as a function of the energy of the incoming electron:
\begin{equation}
    T(E) = \mathrm{Tr}\left( \Gamma_{S} G^{r} \Gamma_{D} G^{a} \right).
\end{equation}

In the Landauer--B\"uttiker formalism \cite{Datta2005} we can use the value of $T(E)$ to evaluate the current in the channel as a function of applied bias:
\begin{equation}
    I = \frac{2e^2}{h} \int_{-\infty}^{\infty} T(E) \left[ f_S(E) - f_D(E) \right] dE
    \label{eqn:landauer}
\end{equation}
where $e$ is the charge of an electron, $h$ is Planck's constant, and $f_i(E)$ is the Fermi--Dirac distribution for contact $i$
\begin{equation}
    f_i(E) = \left[ \exp\left(\frac{E - \mu_0 - e V_i / 2}{k_B T}\right) + 1 \right]^{-1}
\end{equation}
where $\mu_0$ is the Fermi level, $k_B$ is Boltzmann's constant, and $T$ is the temperature.
The junction is symmetrically biased such that $V_S = -V_D$.
Eqn.~\ref{eqn:landauer} could also be used to determine the junction resistance from the gradient of the linear I--V response at low bias.
However the computational cost can be reduced by working in the limit $V \rightarrow 0$ where we can use the zero-bias conductance formula
\begin{equation}
    G = -\frac{2e^2}{h} \int_{-\infty}^{\infty} T(E)\,\frac{\partial f_0(E)}{\partial E} \ dE
\end{equation}
where $f_0(E)$ is the equilibrium Fermi-Dirac distribution function
\begin{equation} 
    \label{eq:equil_fermi_function_derivative}
    \frac{\partial f_0(E)}{\partial E} = -\frac{1}{4 k_B T}\, \mathrm{sech}^2 \left(\frac{E - \mu_0}{2 k_B T}\right),
\end{equation}

A further optimisation is obtained by taking the zero-temperature limit where Eqn.~\ref{eq:equil_fermi_function_derivative} becomes a delta function centered at $\mu_0$.
This gives us an expression for the resistance which only requires the evaluation of the transmission function at a single energy:
\begin{equation}
    R_N = \frac{2}{G_0 T(E = \mu_0)}
    \label{eqn:rn}
\end{equation}
where $G_0 = 2e^2 / h$ is the conductance quantum.

We report the resistance-area $R_NA$ given by the product of the normal resistance (calculated with Eqn.~\ref{eqn:rn}) and the area of the simulation cell transverse to the conduction direction.
We choose to calculate the resistance-area as it is commonly measured in experiment and can be calculated assuming normal state conduction.
With the Ambegaokar--Baratoff relation
\begin{equation}
    I_C R_N = \frac{\pi \Delta}{2e} \tanh \left( \frac{\Delta}{2 k_B T} \right)
\end{equation}
we are then able to link the resistance in the normal state with the critical current of the device when it is superconducting.\cite{Ambegaokar1966}

Calculations of the current and resistance (with Eqns.~\ref{eqn:landauer} and \ref{eqn:rn} respectively) depends on the value of the Fermi level $\mu_0$.
To estimate $\mu_0$ we fit our simulation to an experimental value of the resistance-area.
For a reference junction (with typical thickness, density, and stoichiometry) we calculate the resistance-area for a range of energies.
The Fermi level is then found by matching the calculated resistance-area with a representative experimental\cite{Aref2014} resistance-area of 600~$\Omega\ \mu$m$^2$.
For our data set this gives a value of $\mu_0 = 1.35$~eV.
In Sec.~\ref{sec:results} we are limited to a discussion of qualitative trends only as variation in $\mu_0$, which would occur if a different junction or experimental value was chosen as a reference point, leads to an offset in the calculated resistances for the junction models.

We can also calculate electronic properties which vary spatially within the junction structure.
The charge density in three-dimensions is given by 
\begin{equation}
    n(x,y,z) = -\frac{i}{2 \pi a_x a_y a_z} \mathrm{diag}\left( G^n(E) \right)
    \label{eqn:charge_xyz}
\end{equation}
where the electron Green's function
\begin{equation}
    G^n(E) = G^r(E) \Sigma^{\mathrm{in}}(E) G^r(E)^{\dagger}
\end{equation}
and
\begin{equation}
    \Sigma^{\mathrm{in}}(E) = \Sigma_S(E) f_S(E) + \Sigma_D(E) f_D(E).
\end{equation}

The current flowing between two points can be determined from the element-wise product of $H$ and $G^n$:
\begin{equation}
    J(\B{r}, \B{r^{\prime}}, E) = \frac{e}{h}\, \mathrm{Im}\left[H(E) \circ G^n(E)\right].
\end{equation}
The net current in a particular direction is then calculated from the difference between pairs of points and normalised by the area of the discretisation in the other two directions. For example
\begin{equation}
    J_z(x, y, z; E) = \frac{1}{a_x a_y} \left[ J(\B{r}, \B{r^{\prime}}, E)  - J(\B{r^{\prime}}, \B{r}, E) \right]
    \label{eqn:current_xyz}
\end{equation}
where $\B{r} = (x, y, z)$ and $\B{r^{\prime}} = (x, y, z + a_z)$.
The expressions for $J_x$ and $J_y$ are constructed similarly.
It is worth noting that the equations presented here are entirely general to any one-, two-, or three-dimensional transport system that is well described by a nearest neighbour finite difference model.

\section{Results}
\label{sec:results}

\subsection{Current--voltage response}

\begin{figure}
    \includegraphics{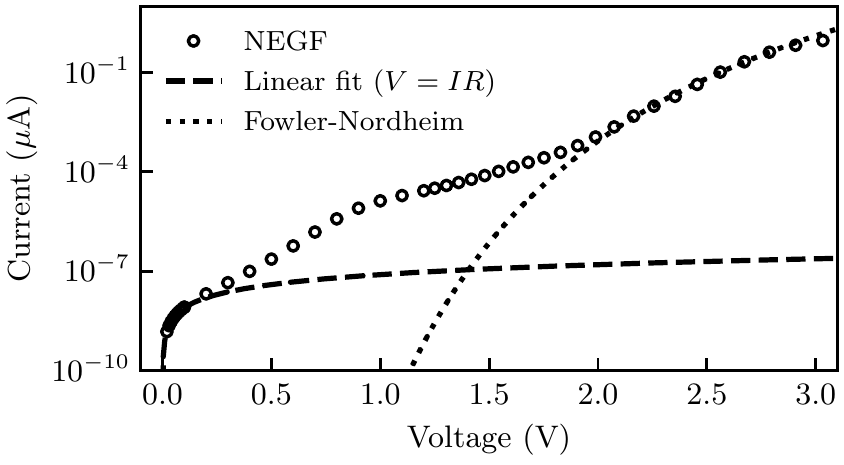}
    \caption{
        The current-voltage response of an \aloxal\ junction model calculated with our NEGF model.
        The material properties of the oxide ($d = 20~$\AA, $\rho = 0.8$, and $\gamma = 1.25$) are close to the mean of experimental reports.\cite{Tan2005,Sullivan1998a,Zeng2015a}
        At low bias a linear response is observed, while agreement with the Fowler--Nordheim model is seen at high bias.
    }
    \label{fig:fowler_nordheim_plot}
\end{figure}

The response of a tunnel junction to an applied bias is expected to be linear when the bias is close to zero and to become non-linear as the bias is increased.
When a sufficiently large bias is applied the Fowler--Nordheim tunnelling theory can be used to describe the response \cite{Fowler1928}.
The current--voltage relationship for an atomistic junction model, calculated in three-dimensions with Eqn.~\ref{eqn:landauer}, is shown in Fig.~\ref{fig:fowler_nordheim_plot}.
We observe a linear response at low bias and find that the behaviour is well described by the Fowler--Nordheim tunnelling model above an applied voltage of approximately 2~V.
It should be noted that this is well above the typical experimental breakdown voltage for junctions and is used here simply to benchmark the technique.

The current in the Fowler--Nordheim model is given by \cite{Hartman1964,Lenzlinger1969,Kaltenbrunner2011}
\begin{equation}
    I(V) = \frac{\alpha A a \beta^2 V^2}{\phi}\ \exp \left( -\frac{b\phi^{3/2}}{\beta V} \right)
    \label{eqn:fowler_nordheim_equation}
\end{equation}
where $\alpha$ is a scaling factor related to the proportion of the barrier which participates in tunnelling via field emission, $A$ is the cross-sectional area of the device, $\phi$ is the work function and $\beta$ is the inverse of the barrier thickness $d$.
The quantities $a$ and $b$ are the Fowler--Nordheim constants which are given by
\begin{equation}
    a = \frac{e^3}{8\pi h} \quad \mbox{and} \quad b = \frac{8\pi}{3} \frac{\sqrt{2 \widetilde{m}}}{eh}
\end{equation}
where $\widetilde{m}$ is the effective mass of the electron in the oxide.
An estimate of the work function $\phi$ can be found by fitting the calculated current-voltage data with Eqn.~\ref{eqn:fowler_nordheim_equation}.

The effective mass used during the fitting process affects the calculated value of the work function.
An effective mass of $\widetilde{m} = 0.4\ m_e$ estimated from band structure calculations\cite{Perevalov2009,Xu1991} for crystalline \corundum\ yields a value for the work function of $\phi = 2.4$~eV.
Alternatively, direct measurement of aluminium oxide barriers\cite{Rippard2002a} gives an estimate of $\widetilde{m} = 0.75\ m_e$ leading to a value of $\phi = 2.0$~eV.
Both values are close to the commonly quoted barrier height\cite{Snow1996,Jeurgens2002,Gloos2003,Hasnaoui2005} of 2~eV.

\subsection{Effect of oxide morphology on resistance-area}
\label{sec:rna}

\begin{figure}[b]
    \includegraphics{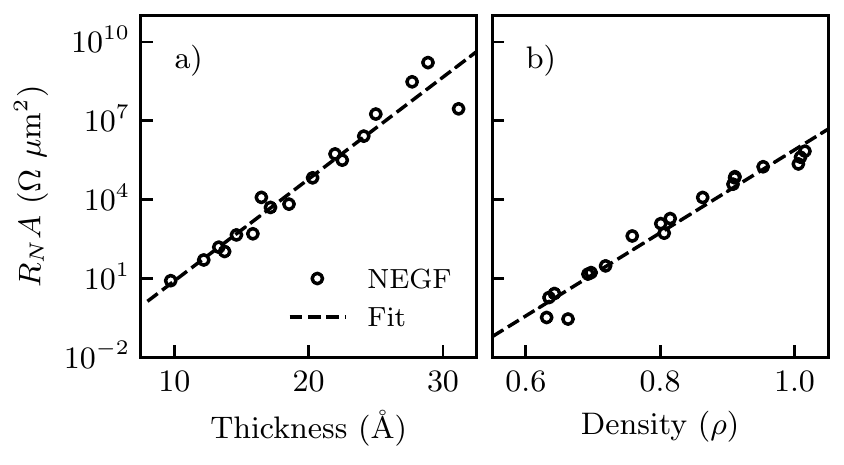}
    \caption{
        The calculated resistance-area of the junctions is exponentially dependent on both a) the thickness of the tunnelling barrier and b) the density of the barrier oxide.
        Linear fits to the log of the resistance-area data are calculated with MATLAB.
    }
    \label{fig:resistances_thickness_density}
\end{figure}

The resistance-area is calculated with Eqn.~\ref{eqn:rn} for each junction in the three data sets described in Table~\ref{tab:datasets}.
The resistance-area as a function of oxide thickness is shown in Fig.~\ref{fig:resistances_thickness_density}a.
A linear fit to the log of the resistance-area data is calculated using MATLAB.
This data set consists of 18 junctions with approximate thicknesses between 10 and 30~\AA\ and densities in the narrow range $\rho$ = 0.77--0.87.
The exponential increase in the resistance-area with barrier thickness is in agreement with experimental observations\cite{Dorneles2003} and an exponential reduction in the tunnelling probability.

Fig.~\ref{fig:resistances_thickness_density}b shows the relationship between the density of the barrier oxide and the resistance-area.
Here we observe that the resistance of the junction is also exponentially related to the oxide density.
We note that each junction in this second data set has a similar thickness ($d = 16 \pm 1$~\AA).
To the authors' knowledge, no systematic studies exist investigating the relationship between the junction resistance and the oxide density.
\citet{Sullivan1998a} report that oxides manufactured with an O$_2$ plasma deposition process are of higher density ($\rho$~=~0.8) when Al is evaporated simulataneously and lower density ($\rho$~=~0.6--0.7) when the substrate is exposed only to the plasma.
From Fig.~\ref{fig:resistances_thickness_density}b we can estimate that this variation in the density would correspond to change in the resistance-area of 1--2 orders of magnitude.

\begin{figure}
    \includegraphics{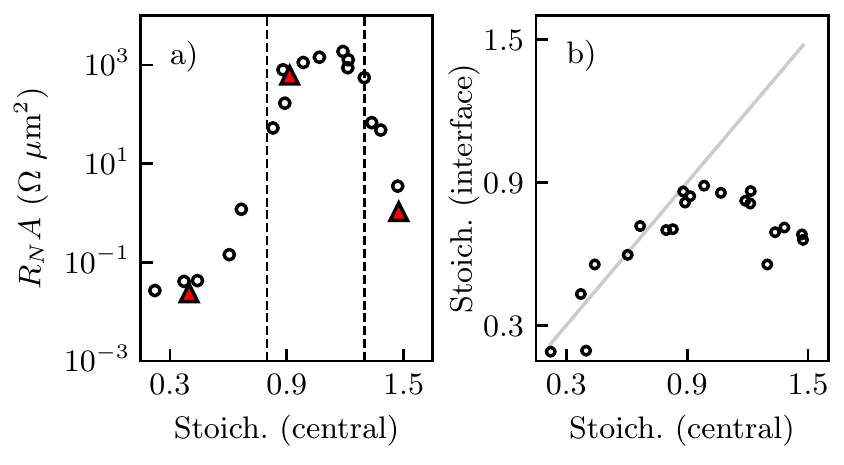}
    \caption{
        a) The calculated resistance-area varies over several orders of magnitude as the oxide goes from an oxygen deficient to an oxygen rich configuration.
        Charge and current densities are presented in Fig.~\ref{fig:charge_dens_stoichiometries} for the structures corresponding to the three red triangles.
        b) The stoichiometry in both the centre of the oxide barrier and in the interfaces between the oxide and the aluminium contacts are plotted.
        The grey line illustrates deviation from uniform stoichiometry across the width of the oxide.
        The central and interfacial regions are defined in Fig.~\ref{fig:example_junction}.
    }
    \label{fig:resistances_stoichiometry}
\end{figure}

Resistance-area data are presented in Fig.~\ref{fig:resistances_stoichiometry}a for a range of oxide stoichiometries .
Between $\gamma$ = 0.9 and $\gamma \simeq$ 1.2 the resistance-area is approximately constant.
This range is comparable to reported experimental values for oxide stoichiometry of $\gamma$ = 0.8--1.2 (depending on fabrication conditions).\cite{Tan2005}
A significant drop in the resistance is seen for values of $\gamma$ outside this region.
At low stoichiometries ($\gamma <$~0.9) this is due to oxygen deficiency in the junctions creating metallic channels which dominate conduction and lead to a decreased resistance.

Fig.~\ref{fig:resistances_stoichiometry}b helps to explain the decreased resistance at higher stoichiometries ($\gamma > 1.2$).
We define the stoichiometry in the interfaces as $\gamma_{\mathrm{interface}}$ and plot how this changes as a function of the stoichiometry in the centre of the barrier $\gamma$. 
The central and interfacial regions are defined in Fig.~\ref{fig:example_junction}.
We observe that $\gamma_{\mathrm{interface}}$ begins to decrease at higher values of $\gamma$.
This implies that there is more aluminium than oxygen in the interfacial regions between the contacts and the oxide barrier.
Constraints used during the preparation of the structure such as requiring a density of $\rho = 0.8$ and limiting the motion of atoms during the optimisation are forcing the system away from thermodynamic equilibrium and may drive this variation in stoichiometry across the oxide.
The oxygen deficient interfaces are more conductive and cause a decrease in the effective thickness of the tunnelling barrier leading to the observed decrease in the resistance-area product.
We note that while high conductance is observed at both low and high stoichiometries, the transport in the high stoichiometry region is still in the tunnelling regime where $T(E) < 1$.

\subsection{Charge and current density}

\begin{figure}
    \includegraphics[width=1\columnwidth]{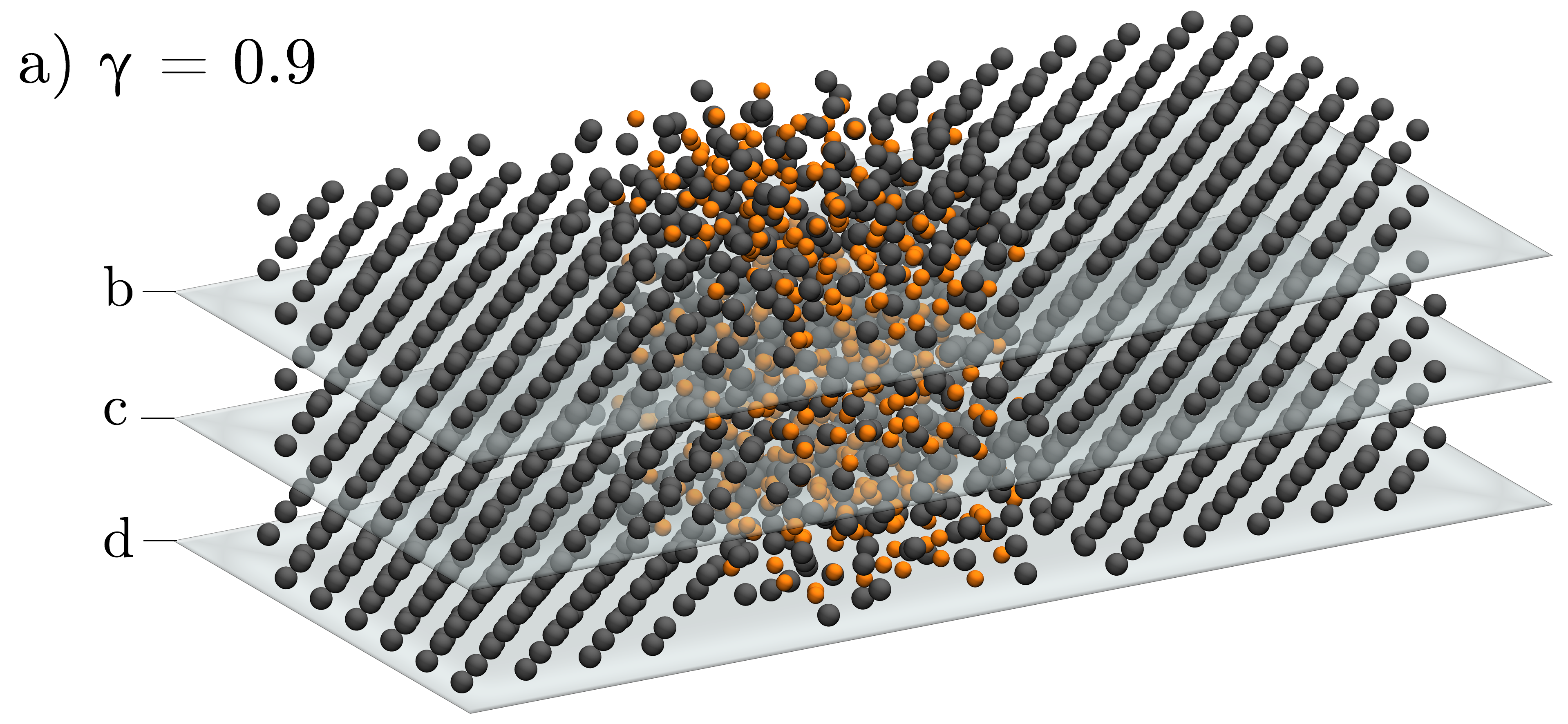}
    \begin{flushright}
        \includegraphics{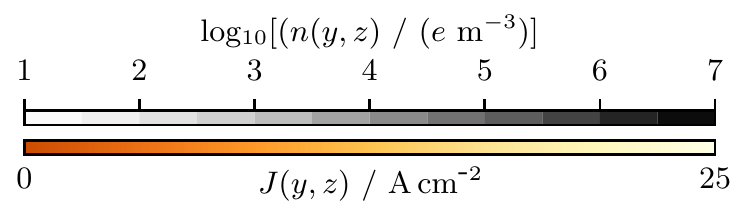}
    \end{flushright}
    \vspace{-12pt}
    \includegraphics{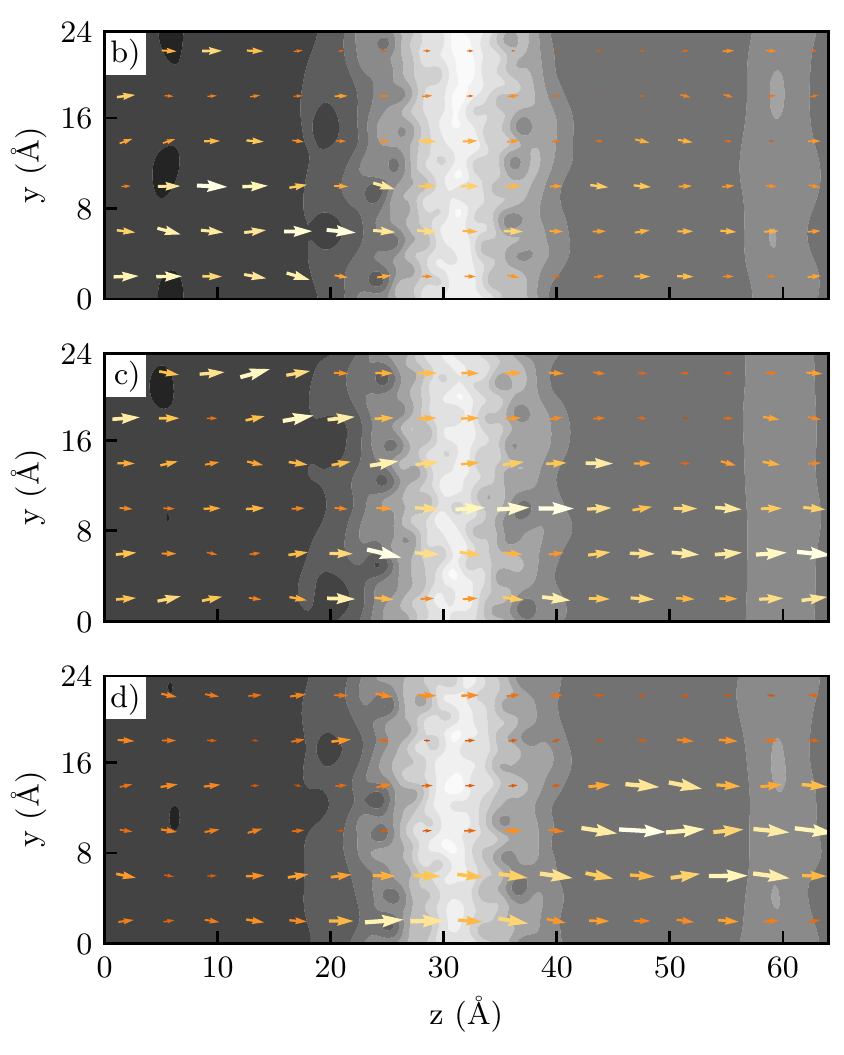}
    \caption{
        The calculated charge density $n(y,z)$ and current density $J(y,z)$ (for an applied bias of 50~mV) are shown for a junction model with a stoichiometry of $\gamma$ = 0.9.
        This structure has a material density of $\rho$ = 0.8 and a barrier thickness of $d$ = 15~\AA.
        The subfigures b)--d) correspond to different positions along the $x$-axis as shown in a).
    }
    \label{fig:charge_dens_s090}
\end{figure}

To better understand how conductance changes as a function of stoichiometry, we calculate the charge density and current density in three dimensions for junction models with stoichiometries of $\gamma = $ 0.3, 0.9 and 1.5 (corresponding to the three red triangles in Fig.~\ref{fig:resistances_stoichiometry}).
These properties were computed with Eqns.~\ref{eqn:charge_xyz} and \ref{eqn:current_xyz} at an applied bias of 50~mV.
In Fig.~\ref{fig:charge_dens_s090} we plot $n(y,z)$ and $J(y,z)$ for a junction in the tunnelling regime ($\gamma$ = 0.9) for three planes at different positions along the $x$-axis.
Lighter regions with lower charge density are associated with the presence of oxygen.
The disorder in the atomic structure of the oxide can be observed in the contours of the charge density in the barrier region.
The current density varies as a function of $x$ and $y$ with regions of higher current around the centre of Fig.~\ref{fig:charge_dens_s090}c and on the bottom of Fig.~\ref{fig:charge_dens_s090}d.

Variation in the physical thickness of the oxide layer has been observed directly in microscopy studies where it is estimated that less than 10\% of the total barrier area dominates the tunnelling of electrons.\cite{Zeng2015a}
Our results demonstrate that the effective width of the tunnelling barrier can be affected by small local differences in the density of aluminium and oxygen atoms.
This is evident even in our junction models with minimal variation in physical thickness across the structure.

\begin{figure}
    \begin{flushright}
        \includegraphics{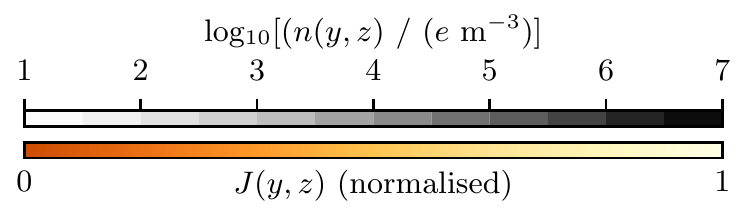}
    \end{flushright}
    \vspace{-12pt}
    \includegraphics{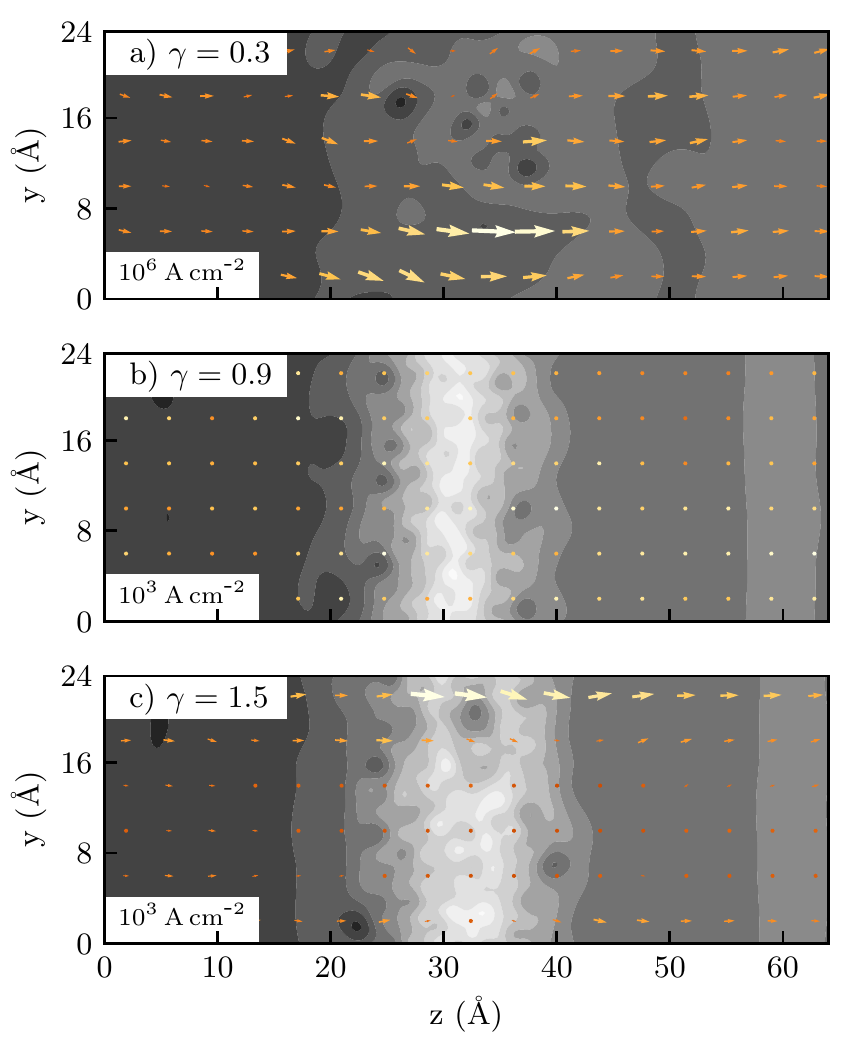}
    \caption{
        The calculated charge density $n(y,z)$ and current density $J(y,z)$ (for an applied bias of 50~mV) are shown for junction models of increasing stoichiometry; a) $\gamma$ = 0.3, b) $\gamma$ = 0.9, and c) $\gamma$ = 1.5.
        Each structure has a material density of $\rho$ = 0.8 and a barrier thickness of $d$ = 15~\AA.
        The current density is normalised within each subfigure by the value shown in the lower left corner.
    }
    \label{fig:charge_dens_stoichiometries}

    \vspace{36pt}

\end{figure}

A comparison of the calculated current for the various stoichiometries is shown in Fig.~\ref{fig:charge_dens_stoichiometries}.
The charge density contours in Fig.~\ref{fig:charge_dens_stoichiometries}a ($\gamma$ = 0.3) show a significantly weaker suppression of the current than is evident in the insulating $\gamma$ = 0.9 junction.
The low stoichiometry structure ($\gamma$ = 0.3) contains small regions of aluminium oxide that do not span the entire lateral width of the junction model, leaving metallic channels through which the majority of the current flows.

Figures~\ref{fig:charge_dens_stoichiometries}b and c show the charge and current density for the higher stoichiometry structures ($\gamma$ = 0.9 and 1.5).
It is important to note here that the arrows depicting the current density are 10$^3$ times smaller than those in Fig.~\ref{fig:charge_dens_stoichiometries}a.
Fig.~\ref{fig:charge_dens_stoichiometries}b ($\gamma$ = 0.9) corresponds to a junction in the fully insulating regime, while a path of higher current density can be seen at the top of Fig.~\ref{fig:charge_dens_stoichiometries}c ($\gamma$ = 1.5).
We believe this arises because structures at stoichiometries higher than the experimentally observed values are not in thermodynamic equilibrium.
Oxygen deficiency in the Al/AlO$_\mathrm{x}$ interfaces creates areas where the insulating barrier is thinner and electrons can more easily tunnel through the oxide.
The current densities presented in Fig.~\ref{fig:charge_dens_stoichiometries} allow us to understand the drop in the calculated resistance-area values (at both low and high stoichiometries) in Fig.~\ref{fig:resistances_stoichiometry}.

\section{Conclusions}

Fine control of the critical current is highly desirable for creating addressable qubits when fabricating devices containing tens or hundreds of Josephson junctions.
In this work we study the interplay between the internal structure of the oxide and its electrical characteristics using a three-dimensional description of the junction.
The material properties of the oxide layer in the \aloxal\ junction are found to affect the calculated resistance-area product.
We observe the exponential dependence between the thickness of the oxide barrier and the junction resistance as expected.
The junction resistance also changes with the stoichiometry of the barrier with conduction in highly sub-stoichiometric structures being dominated by metallic conduction channels.
Additionally we find that the junction resistance is exponentially dependent on the oxide density.

To study how the electronic characteristics change due to local atomic structure we calculate the charge density and current density.
In highly oxygen deficient structures conduction is dominated by metallic channels.
However, even with more oxygen present, particular paths through the oxide contribute more to the current flow. 
This non-uniformity of the current distribution has important consequences for the influence of charged defects within the amorphous structure. 
Defects near dominant conduction paths are more likely to couple strongly to the current, contributing to the noise in the critical current $I_C$.

Despite their widespread usage, \aloxal\ junctions suffer universally from noise caused by two-level systems whose exact physical origin is an ongoing topic of interest.\cite{Muller2017}
Magnetic surface spins,\cite{DeGraaf2017} delocalised atoms,\cite{DuBois2013,DuBois2014} and many other models have been proposed to explain the observed noise.\cite{Cole2010}
Understanding the physical origin of two-level defects and their impact on the electrical properties of junctions is key in achieving improvements and consistency in fabrication.
The present work provides a framework for testing TLS models and developing a better understanding of how the performance of a junction in a circuit relates to its atomic structure.

We have developed a computational approach for determining the electrical characteristics of \aloxal\ junction models based on their atomistic structure. 
Using this technique allows us to study the role of junction morphology and composition in determining junction performance.
An understanding of the exponential dependence of the junction resistance on barrier thickness and oxide density can be reached using relatively simple models.
However, the relationship between the atomic structure and flow of current through the junction can only be fully understood with a complete three-dimensional treatment of the problem.
Developing computational modelling tools for atomistic simulation of electronic devices at the nanoscale will prove invaluable in optimising their fabrication, leading to more reliable and reproducible nanoelectronics.

\begin{acknowledgments}
    The authors acknowledge support of the Australian Research Council through grants DP140100375, CE170100026 (MJC), and CE170100039 (JSS).
    The authors also acknowledge useful discussions with J. Gale.
    This research was undertaken with the assistance of resources from the National Computational Infrastructure (NCI), which is supported by the Australian Government.
\end{acknowledgments}

\bibliography{transport}

\clearpage

\appendix
\section{Notes on numerical approximations}
\label{sec:potential}

\subsection{Truncation of the Coulombic potential}

In order to calculate the electrostatic potential $V(x,y,z)$ inside the junction structures we use the Ewald summation method.
This is a standard method for computing the electrostatic energy of a particular configuration of charges in a periodic system.\cite{Ewald1921}
We implement a version of Ewald summation in which a modified version of the short range real space interaction is used.
This is necessary because the finite difference approach used in our NEGF calculations becomes a poor approximation when confronted with the divergences arising from the Coulombic potential close to a charged particle.
To account for this we replace the Coulombic potential for short-range interactions with a potential of a Gaussian form.
The junction potential is calculated on an evenly spaced three-dimensional grid using the coordinates and charges obtained from the molecular dynamics calculation.

We define a radius $r_c$ inside which the Gaussian-like description of the potential will be used and write down the function $h(r)$ which combines the Coulombic and Gaussian components.
\begin{align}
    h(r) &=
        \begin{cases}
            \frac{1}{r_c} \exp \left[\frac{1}{2} - \frac{r^2}{2{r_c}^2} \right] & \quad \left| r \right| < r_c \\
            \frac{1}{r} & \quad \mathrm{otherwise.}
        \end{cases}
\end{align}

The different potential profiles are shown in Fig.~\ref{fig:coulomb_vs_gaussian_potential} where the red points indicate the potential used in our calculations. 
The energy scale is characteristic of the atomic sites in a our calculations where the magnitude of the potential of the order of tens of electron-volts.
As we are interested in energies close to the Fermi energy $\mu_0$ ($\sim$~1~eV), the application of the truncation still allows for the atomic structure inside the junction models to be reflected in the calculated electronic properties.
By using the Gaussian-like potential to describe the short range interactions we ensure that the potential varies smoothly throughout the junction structure and avoid the numerical instabilities of the Coulombic divergences.

\subsection{Finite difference order}

\begin{figure*}
    \centering
    \includegraphics{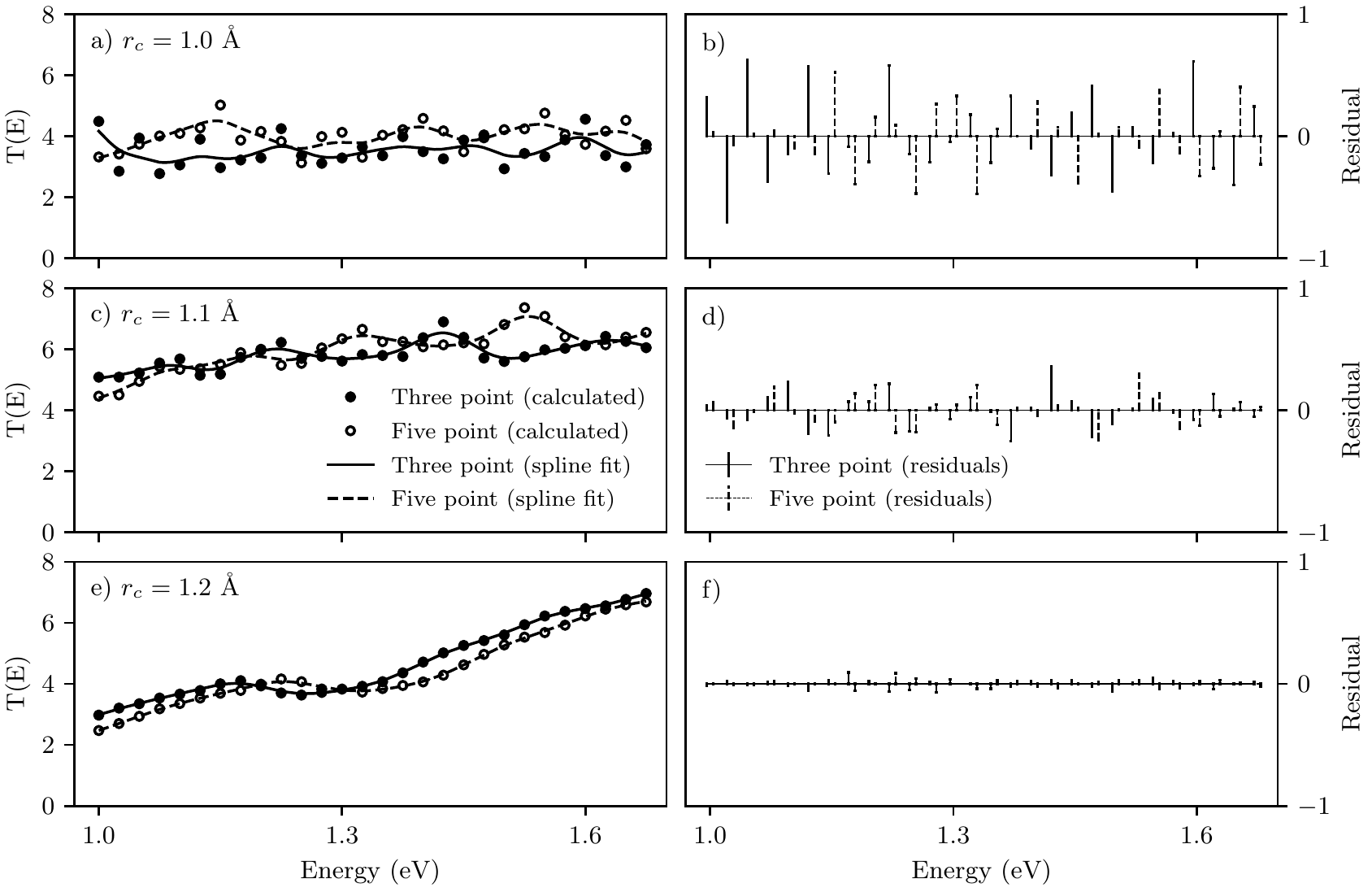}
    \caption
    {
        The three-dimensional transmission function calculated for $r_c$ = 1.0, 1.1, and 1.2~\AA.
        Smoothing spline fits calculated with MATLAB are also shown.
        The  residuals show the behaviour of the transmission function becoming smoother as the radius $r_c$ is decreased.
        A value of $r_c$ = 1.2~\AA\ was found to produce a smooth and continuous transmission function.
    }
    \label{fig:transmission_as_fn_of_rc}
\end{figure*}

To choose an appropriate value for $r_c$ the transmission was calculated in three-dimensions for a range of radii using both three- and five-point finite difference approximations.
The transmission is plotted in the left hand panels of Fig.~\ref{fig:transmission_as_fn_of_rc} (shown on pg.~\pageref{fig:transmission_as_fn_of_rc}) for values of $r_c$ = 1.0, 1.1, and 1.2~\AA\ along with smoothing splines fitted with MATLAB.
On the right hand side the residuals are plotted showing the difference between the calculated transmission and the fitted spline.
As the radial truncation increases the Coulombic divergences are smoothed out which in turns affects the stability of the calculated transmission.
The behaviour of the residuals is more dependent on the value of $r_c$ than the order of the finite difference approximation.

\begin{figure}[b]
    \centering
    \includegraphics{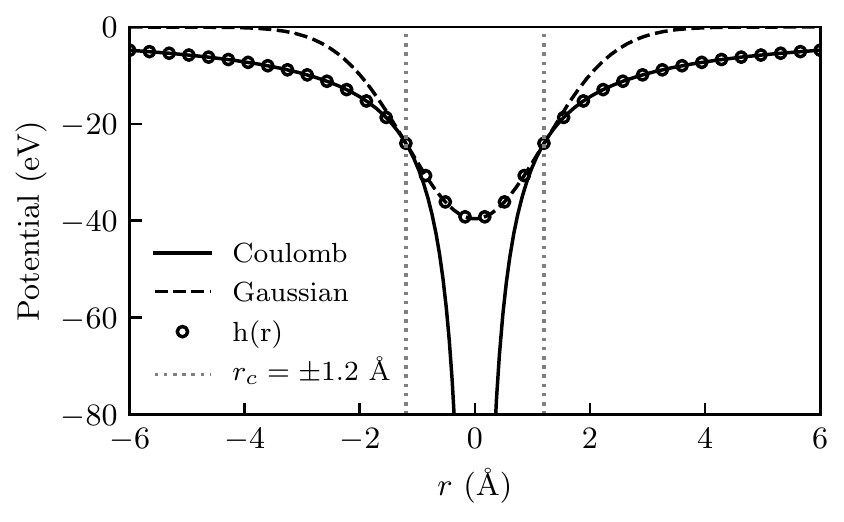}
    
    \caption
    {
        The Coulombic and Gaussian potentials are shown for a point charge with $q = -2|e|$.
        The two potentials are matched to have an equal value and gradient at $r = \pm r_c$.
        The function $h(r)$ is also shown where interactions at $r < \pm r_c$. are described by the Gaussian potential, while the Coulombic potential is used for $r > r_c$.
        The markers are separated by a distance of 1/3~\AA\ which is representative of the discretisation used.
    }
    \label{fig:coulomb_vs_gaussian_potential}

    \vspace{12pt}

    \includegraphics{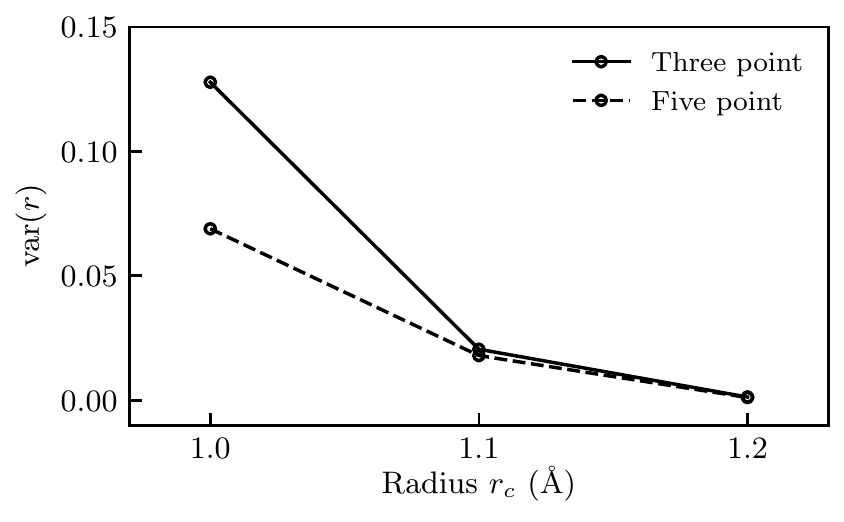}
    \caption{
        The variance of the residuals $r$ in Fig.~\ref{fig:transmission_as_fn_of_rc} is plotted as a function of the truncation radius $r_c$ for three- and five-point finite difference approximations.
    }
    \label{fig:fd_residual_var}
\end{figure}

To obtain a single metric for the smoothness of the transmission calculation we calculate the variance of the residuals.
Fig.~\ref{fig:fd_residual_var} shows the decrease in the variance of the calculated residuals $\mathrm{var}(r)$ as $r_c$ increases and also highlights that the choice of $r_c$ affects numerical accuracy more than changing the finite difference approximation.
With the view to include as much of the physics around the atomic sites as possible we use a value of $r_c$ = 1.2~\AA\ for the remainder of the work.

\end{document}